\documentclass[%
aps,
prd, 
rsi,%
 amsmath,amssymb,
preprint,
author-year,%
author-numerical,%
superscriptaddress, 
secnumarabic, 
nobibnotes, 
citeautoscript
]{revtex4-1}
\usepackage{graphicx}
\usepackage{dcolumn}
\usepackage{bm}
\usepackage[caption=false]{subfig}
\usepackage{diagbox}
\usepackage{color}
\usepackage{mathrsfs}
\usepackage{booktabs}
\usepackage[normalem]{ulem}

\usepackage{hyperref}

\def\be{\begin{equation}}
\def\ee{\end{equation}}

\def\begineqn{\begin{equation*}}
\def\endeqn{\end{equation*}}

\def\beginar{\begin{eqnarray}}
\def\endar{\end{eqnarray}}
\def\beginarn{\begin{eqnarray*}}
\def\endarn{\end{eqnarray*}}

\def\lb{\left ( }
\def\rb{\right ) }
\def\lsq{\left [ }
\def\rsq{\right ] }

\def\ep{\epsilon}

\def\Rat{\widetilde{Ra}}
\def\Qt{\widetilde{Q}}

\def\ubp{\mathbf{u}}
\def\bbp{\mathbf{b}}

\def\mBb{\overline{\bf B}}

\def\mth{\overline{\Theta}}
\def\pth{\vartheta}

\def\pbz{b_z}

\def\pjz{j_z}

\def\dst{{\partial_t}}

\def\dz{{\partial_Z}}
\def\dzt{{\partial^2_Z}}

\def\md{D^\perp_t}

\def\lp{{\nabla_\perp^2}}

\begin{document}


\title[Quenching rapidly rotating turbulence]{Magnetic quenching of the inverse cascade in rapidly rotating convective turbulence}

\author{Stefano Maffei}
\affiliation{ 
Department of Physics, University of Colorado, Boulder, Colorado 80309, USA
}%
\author{Michael A. Calkins}%
\affiliation{ 
Department of Physics, University of Colorado, Boulder, Colorado 80309, USA
}%

\author{Keith Julien}
\affiliation{%
Department of Applied Mathematics, University of Colorado, Boulder, Colorado 80309, USA
}%

\author{Philippe D. Marti}
\affiliation{ 
Center for Climate System Modeling, ETH Z\"{u}rich, CH-8092, Switzerland
}%
\affiliation{ 
Department of Earth Sciences, ETH Z\"{u}rich, CH-8092, Switzerland
}%

\date{\today}

\begin{abstract}

We present results from an asymptotic magnetohydrodynamic model that is suited for studying the rapidly rotating, low viscosity regime typical of the electrically conducting fluid interiors of planets and stars. We show that the presence of sufficiently strong magnetic fields  prevents the formation of large-scale vortices and saturates the inverse cascade at a finite length-scale. This saturation corresponds to an equilibrated state in which the energetics of the depth-averaged flows are characterized by a balance of convective power input and ohmic dissipation. A quantitative criteria delineating the transition between finite-size flows and domain-filling (large-scale) vortices in electrically conducting fluids is found. By making use of the inferred and observed properties of planetary interiors, our results suggest that convection-driven large-scale vortices do not form in the electrically conducting regions of many bodies.


\end{abstract}

\maketitle

The subsurface regions of stars and the fluid cores of planets are typically characterized by rapid rotation, buoyancy-driven convective turbulence, and electromagnetic fields generated by the dynamo mechanism that converts the kinetic energy of fluid motion into electromagnetic energy. The dynamical state of such systems is characterized by several non-dimensional parameters, including the Reynolds number, $Re_H = U H/\nu$, the Ekman number, $E_H = \nu/(2\Omega H^2)$, and the Rossby number, $Ro_H=U/(2\Omega H)$. Here $U$ and $H$ represent a typical speed and length-scale of the flow, $\nu$ is the kinematic viscosity and $\Omega$ is the rotation rate. The Reynolds, Ekman and Rossby numbers represent the relative sizes of inertia to viscous forces, viscous forces to the Coriolis force, and inertia to the Coriolis force, respectively. Rapidly rotating turbulent flows are characterized by $Re_H \gg 1$ and $E_H \ll Ro_H \ll 1$.  An important physical property of electrically conducting fluids is the magnetic Prandtl number, $Pm = \nu/\eta$, where $\eta$ is the magnetic diffusivity. For the Earth's liquid outer core these parameters are estimated to be $Re_H \approx 10^8$, $E_H \approx 10^{-15}$, $Ro_H \approx 10^{-7}$ and $Pm \approx 10^{-6}$  \cite{olson20158}. In contrast, the most extreme direct numerical simulation (DNS) spherical dynamo study to date  \cite{schaeffer_turbulent_2016} used values of $E_H = 5 \times 10^{-8}$ and $Pm = 10^{-1}$, and reached $Re_H \approx 5 \times 10^3$. Although of crucial importance for understanding magnetohydrodynamics (MHD), it is unknown how the results of such DNS studies extrapolate to natural systems. For flows with sufficiently large values of $Re_H$, and subject to a broad variety of forcing mechanisms, rapidly rotating three-dimensional turbulence gives rise to the formation of motions with large lateral scales relative to the forcing scale  \cite{smith1999transfer,favier2014inverse,rubio_upscale_2014,guervilly_large_2014,tL17}. Such flows result from an inverse energy cascade that leads to a net transfer of kinetic energy from the small-scale motions to large-scale motions of the flow. 
For hydrodynamic, non-magnetic turbulence, the ultimate scale at which the inverse cascade ceases is dependent upon the geometry. In a Cartesian geometry with equal horizontal dimensions, the cascade leads to domain-filling large-scale vortices (LSVs)  \cite{guervilly2015generation,tL17,rubio_upscale_2014}. In a spherical geometry, the inverse cascade is halted at the so-called Rhines scale  \cite{pR75}. The Rhines scale represents a dynamical cross-over between eddy dynamics that dominate on small-scales and Rossby wave dynamics that dominate on large-scales; it is thought to control the latitudinal extent of alternating winds in the outer, electrically-insulating fluid regions of giant planet atmospheres \cite{sC17}. 

It has been suggested, based on the results of DNS studies \cite{guervilly2015generation,lin2016precession,guervilly2017large}, that inverse-cascade-generated LSVs might be important for generating large-scale (i.e.~domain-scale) magnetic fields in planets and stars. However, rapid rotation alone is sufficient for generating large-scale magnetic fields \cite{calkins_multiscale_2015}, even for laminar, small-Reynolds-number flows that lack an inverse cascade \cite{aS74,mC16,mC16b}. In the asymptotic limit of rapid rotation,  LSVs are unimportant for the onset of dynamo action \cite{mC16b}. However, the influence of magnetic field on the inverse cascade remains poorly understood due, in large part, to the limited parameter range of previous DNS studies. It is evident that the Rossby number, in particular, is not low enough in many DNS investigations to be applicable to planetary systems; this effect is evident in DNS studies in which LSVs shows a preference for cyclonic circulation (i.e.~in the same direction as the system rotation vector) \cite{guervilly2015generation,favier2014inverse}. For sufficiently small Rossby numbers the LSV consists of a dipolar vortex with no preference for a particular circulation direction \cite{stellmach_approaching_2014}. Thus, lower Rossby number simulations are necessary to better understand the planetary regime.

 It should be noted that rapid rotation is not a requirement for the inverse cascade. Previous work has shown that imposed magnetic fields can also lead to an inverse energy cascade \cite{hossain1991inverse,alexakis2011two,reddy2014strong}. In addition, studies of two-dimensional turbulence have shown that sufficiently strong magnetic fields \citep{tobias2007beta} and Lorentz-force-like forcing terms \citep{seshasayanan2014edge} can disrupt the inverse cascade. Here we find that a similar effect occurs in rapidly rotating, convection-driven  turbulence, which represents a system that is more applicable to the study of planets and stars.

Although previous work has suggested that the presence of magnetic fields can prevent the formation of large-scale structures in rapidly rotating convective systems \citep{guervilly2017large}, no systematic study has been performed to date that fully elucidates the physical mechanism by which a magnetic field influences the inverse cascade. 
In this regard, we explore the problem by utilizing an asymptotically reduced form of the governing equations of MHD  \cite{calkins_multiscale_2015}, that is valid in the geo- and astrophysically relevant limits of $(E_H, Ro_H, Pm) \rightarrow 0$ with $Re_H\gg1$. We consider rotating Rayleigh-B\'{e}nard convection in a horizontally-periodic plane layer of incompressible fluid of depth $H$, with constant vertical gravity and rotation vectors ($\textbf{g} = -g \textbf{e}_z$ and $\mathbf{\Omega}=\Omega \textbf{e}_z$, respectively, where $\textbf{e}_z$ is the vertical unit vector). A constant temperature difference $\Delta \theta$ between the top and bottom boundaries is maintained to drive convective motions. Here we provide only a brief overview of the derivation of the model. Further details can be found in previous work \citep{sprague_numerical_2006,calkins_multiscale_2015,mC16,mC16b,mP18}. We assume that the small convective spatial scale $l$ and the Rayleigh number $Ra_H=g\alpha\Delta\theta H^3/(\kappa \nu)$ ($\alpha$ is the thermal expansion coefficient and $\kappa$ is the thermal diffusivity) scale, respectively, as $E_H^{1/3} H$ and $E_H^{-4/3}$, as informed from linear theory \cite{chandrasekhar1961hydrodynamic}. These scalings ensure scale separation between $l$ and the depth of the layer such that $\ep \equiv l / H = E_H^{1/3} \ll1$, which translates into separation between the small-scale coordinate system $(x,y,z)$ and the domain-scale, vertical coordinate $Z = \ep z$. The equations are separated into mean (averaged over the small horizontal scales) and fluctuating components. We then expand each dependent variable ($f$, say) in a power series of the form $f = f_0 + \ep^{1/2} f_{1/2} + \ep f_1 + \ldots$, take the limit $\ep \rightarrow 0$, and collect terms of equal magnitude in the resulting system of equations. Upon integrating on the small vertical coordinate $z$, we obtain the following system for the asymptotically reduced equations (where the ordering subscripts on the variables have been dropped):

\begin{equation}
\md \zeta - \dz w = \Qt \, \mBb \cdot \nabla_\perp \pjz + \nabla_\perp^2 \zeta, \label{E:vort0}
\end{equation}
\begin{equation}
\md w +  \dz \psi = \frac{\Rat}{Pr}  \pth  + \Qt \, \mBb \cdot \nabla_\perp \pbz + \lp w, \label{E:mom0}
\end{equation}
\begin{equation}
 \md \pth +  w \dz \mth = \frac{1}{Pr} \lp \pth , \label{E:fheat0}
 \end{equation}
\begin{equation}
 \dz \overline{\lb w \pth \rb} =  \frac{1}{Pr} \dzt \mth , \label{E:mheat} 
 \end{equation}
\begin{equation}
  0 = \mBb \cdot \nabla_\perp \zeta+ \lp  \pjz \label{E:finduc1}, 
  \end{equation}
\begin{equation}
  0 = \mBb \cdot \nabla_\perp w + \lp  \pbz \label{E:finduc2} .
\end{equation}
The overbar denotes an average over fast spatiotemporal scales, and we use the notation $\md = \dst + \ubp_\perp\cdot\nabla_\perp$ and $\nabla_\perp = (\partial_x,\partial_y)$. The geostrophic streamfunction (pressure) is denoted by $\psi$ and defined by $\ubp_\perp = -\nabla_\perp\times(\psi\textbf{e}_z)$; $\zeta = \lp \psi$ is the axial vorticity and $w$ is the vertical velocity; $\pth$ and $\mth$ are the fluctuating and horizontally averaged temperature; $\mBb$, $\pbz$ and $\pjz = \textbf{e}_z\cdot\nabla_\perp\times\bbp$ are the mean magnetic field, fluctuating vertical magnetic field and vertical current density, respectively; $\widetilde{Ra}$ and $\widetilde{Q}$ are, respectively, the asymptotically reduced Rayleigh and Chandrasekhar numbers (see below). For the above set of equations, time has been scaled by the small-scale (horizontal) viscous diffusion timescale $l^2/\nu$, the magnetic field has been scaled by the magnitude of the mean magnetic field $\mathcal{B}$, and temperature has been scaled by $\Delta\theta$. We impose a mean, horizontal magnetic field defined by
\begin{equation}
\mBb = \frac{\sqrt{2}}{2}\lsq \left(\cos(\pi Z) - \cos(3\pi Z)\right)\textbf{e}_x - \cos(\pi Z)\textbf{e}_y\rsq,
\end{equation}
that satisfies perfectly conducting electromagnetic boundary conditions. Mean magnetic fields of similar spatial structure are found to be generated by dynamo action near the onset of rotating Rayleigh-B\'{e}nard convection \cite{aS74,stellmach_cartesian_2004,mC16b,guervilly2017large}. Even for strongly forced convection, the mean field appears to retain a spiralling structure \cite{mC16b,guervilly2017large}.
The use of an imposed, rather than self-generated, magnetic field allows for precise control of the field magnitude. We also utilize the quasi-static MHD approximation on the small convective scale, valid for the small values of $Pm$  typical of planetary and stellar interiors. The dynamics are controlled by three non-dimensional parameters: the asymptotically-scaled Chandrasekhar number, $\widetilde{Q}=Q_H E_H^{2/3}$ (where $Q_H = \mathcal{B}^2H^2/(\mu_0\rho\nu\eta)$ and $\mu_0$ is the magnetic permeability of free space);
  the asymptotically-scaled Rayleigh number, $\widetilde{Ra} = Ra_H E_H^{4/3}$; and the thermal Prandtl number, $Pr=\nu/\kappa$. Here $(\widetilde{Ra}, \widetilde{Q})=O(1)$ (or, more specifically, $\widetilde{Ra}, \widetilde{Q}<\epsilon^{-1/2}$ so that buoyancy and the Lorentz force do not enter lower orders of the asymptotic expansion) and we use $Pr=1$ in order to allow for comparison with previous studies. For the Earth's outer core, $E_H=O(10^{-15})$, and the asymptotic model captures dynamically relevant values of $Ra_H = O(10^{20})$ and $Q_H=O(10^{10})$. The boundary conditions for \eqref{E:vort0}-\eqref{E:finduc2} are impenetrable, stress-free, fixed-temperature and perfectly electrically conducting. 
The equations are discretized in the horizontal and vertical dimensions with Fourier series and Chebyshev polynomials, respectively. The horizontal size of the domain of integration is set to 10 times the critical wavelength at the onset of convection in both the $x$ and $y$ directions. The time-stepping is performed with a third order Runge-Kutta scheme \cite{spalart1991spectral}.  Simulations of the hydrodynamical version ($\widetilde{Q}=0$) of \eqref{E:vort0}-\eqref{E:finduc2} have shown excellent quantitative agreement with laboratory experiments and DNS  \cite{stellmach_approaching_2014,plumley2016effects}. 

\begin{figure*}[tbh]
\centering
{\includegraphics[width=\textwidth]{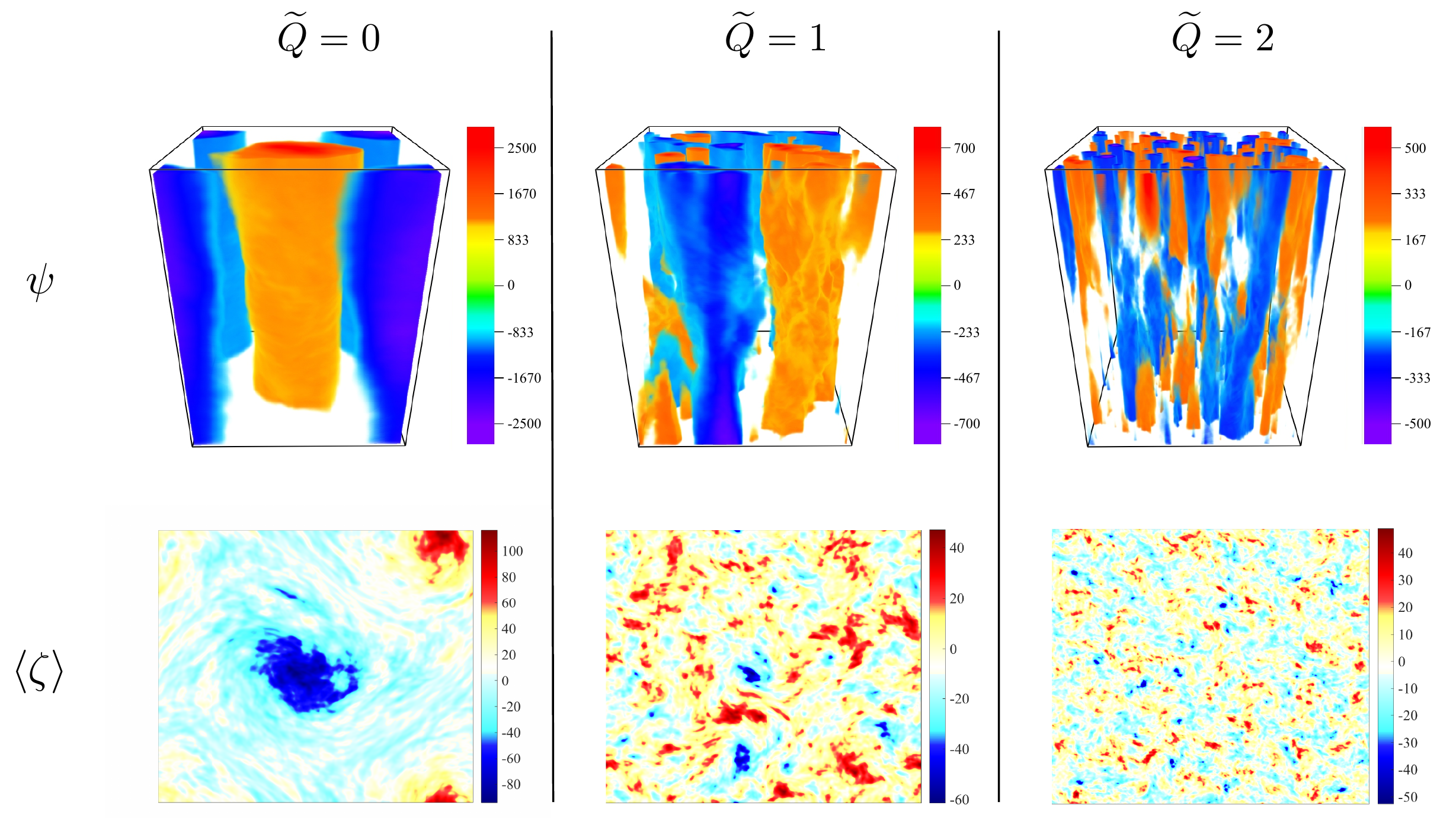}}
\caption{Simulation snapshots. Volumetric renderings of the geostrophic streamfunction ($\psi$, top row) and the depth-averaged axial vorticity ($\left<\zeta\right>$, bottom row). All plots correspond to a reduced Rayleigh number of $\widetilde{Ra}=160$. Three values of the reduced Chandrasekhar number are shown: $\Qt=0$ (first column), $\Qt=1$ (second column) and $\Qt=2$ (third column). }
\label{fig:LSV}
\end{figure*}

Numerical simulations were performed over a broad range of $\widetilde{Q}$ and $\widetilde{Ra}$ (see Supplemental Material at [URL will be inserted by the publisher] for details of the numerical simulations performed in this study), allowing for the investigation of flow regimes ranging from laminar magnetoconvection, up through rapidly rotating magnetoconvective turbulence. For $\widetilde{Q}=0$, $\widetilde{Ra}\ge40$ generates sufficiently turbulent flows that result in the formation of an LSV \cite{rubio_upscale_2014}.  The left panel of Figure \ref{fig:LSV} show instantaneous snapshots of the volume-rendered geostrophic streamfunction (pressure) and vertically integrated axial vorticity for $\widetilde{Q}=0, \Rat=160$. In the rapidly rotating limit considered here, the LSV is dipolar in structure and fills the horizontal extent of the domain such that the (energetically) dominant horizontal wavenumber is the box scale, $k=1$, where $k$ is the modulus of the horizontal wavenumber \textbf{k}. In agreement with previous studies  \cite{julien_statistical_2012,rubio_upscale_2014,guervilly_large_2014}, the baroclinic, convective dynamics is not significantly affected by the presence of the LSV. The central and right panels of Figure \ref{fig:LSV} show the corresponding cases with non-dimensional magnetic field strengths of $\Qt=1$ and $\Qt=2$, respectively, for $\Rat=160$. It is evident that stronger magnetic fields yield a significant reduction in the strength of the horizontal box-scale mode of the depth-averaged motion, to the point that it is no longer visible. 

The formation of the LSV is due to transfer of energy from the convective length-scale (where energy is injected) to the largest scales allowed in the system. This process is described by the spectral evolution equation for the barotropic (vertically integrated, horizontal) kinetic energy $K_{bt}(k)$,
\begin{equation}
\partial_t K_{bt}(k) = T_k + F_k + L_k + D_k.
\label{eqn:dtKbt}
\end{equation}
The four terms on the right-hand side are: (1) the transfer of energy between barotropic modes of wavenumber \textbf{k},
\[
T_k = \sum_{|\textbf{k}|=k}\textrm{Re}\left\{\left<\psi\right>^*_{\textbf{k}} \circ \mathcal{F}_\textbf{k}\left[ J[\left<\psi\right>,\left<\zeta\right>] \right]\right\},
\]
where  $\left<\psi\right>_\textbf{k}$ is the horizontal Fourier transform of the vertically averaged streamfunction $\left<\psi\right>$, the superscript $*$ denotes a complex conjugate, $\mathcal{F}_\textbf{k}[\cdot]$ indicates the horizontal Fourier transform of the argument in square brackets, $J[\cdot]$ is the Jacobian differential operator acting on the arguments in square brackets, the symbol $\circ$ indicates a Hadamard (element-wise) product  (both $\left<\psi\right>_\textbf{k}$ and $\mathcal{F}_\textbf{k}\left[ J[\left<\psi\right>,\left<\zeta\right>]\right]$ are two-dimensional matrices), $\textrm{Re}\left\{\cdot\right\}$ is the real part of the argument in curly brackets and the sum is taken over all horizontal wavenumbers; (2) the transfer of energy between the barotropic and baroclinic (convective) modes 
\[
F_k = \sum_{|\textbf{k}|=k}\textrm{Re}\left\{\left<\psi\right>^*_{\textbf{k}} \circ\mathcal{F}_\textbf{k}\left[ \left<J[\psi',\zeta']\right> \right] \right\};
\] 
(3) the transfer of energy to the barotropic mode from the baroclinic magnetic field
\[
L_k = - \sum_{|\textbf{k}|=k}\widetilde{Q}\textrm{Re}\left\{\left<\psi\right>^*_{\textbf{k}} \circ\mathcal{F}_\textbf{k}\left[ \left<\overline{\textbf{B}}\cdot\nabla j'_z\right>\right] \right\};
\] 
and (4) the viscous dissipation of the barotropic mode
\[
D_k = \sum_{|\textbf{k}|=k}\textrm{Re}\left\{|\textbf{k}|^2 \left<\psi\right>^*_{\textbf{k}}\circ\left<\zeta\right>_{\textbf{k}} \right\}= -2 k^2 K_{bt}(k) .
\]
With the above definitions, positive (negative) values of $T_k$ and $F_k$ indicate energy is being transferred to (from) the barotropic mode $k$ from the interaction of all the other modes. Both $L_k$ and $D_k$ are negative-definite.

\begin{figure*}[tbh]
\centering
{\includegraphics[height=7cm]{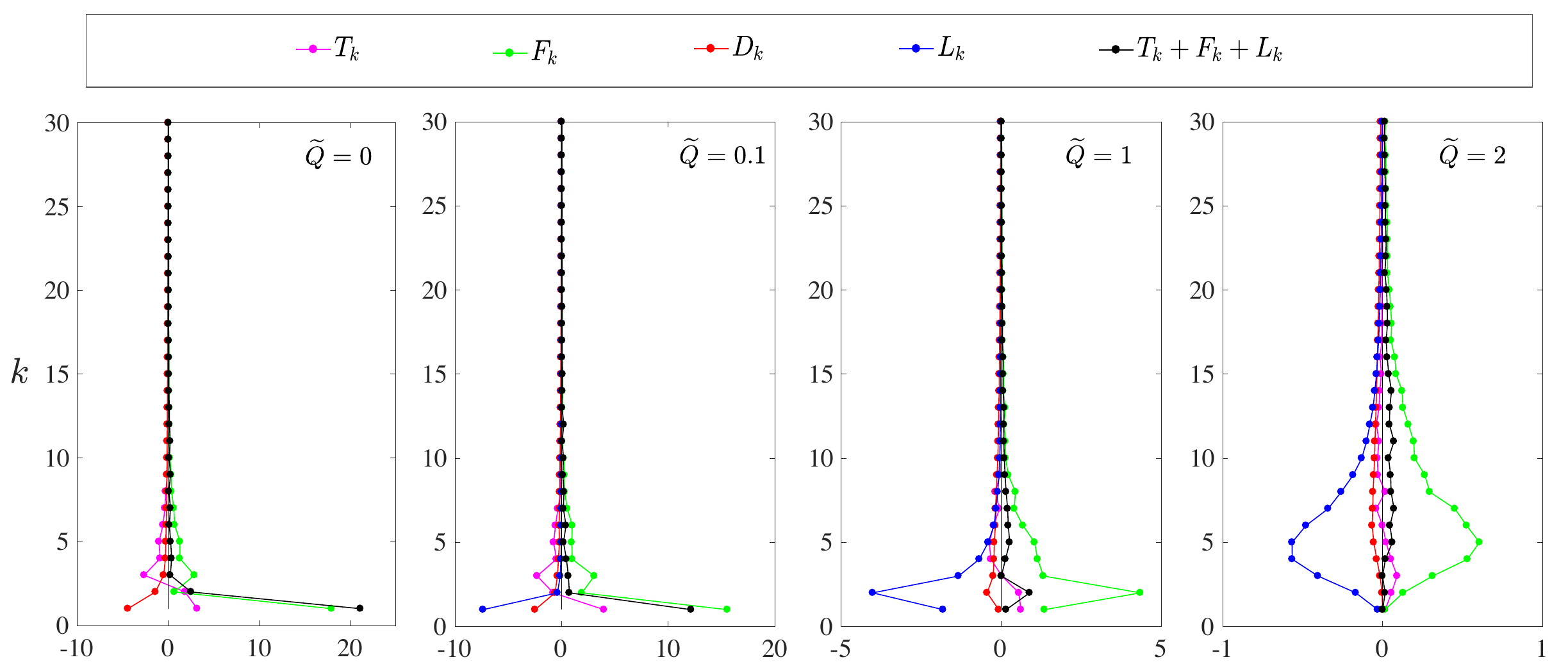}}
\caption{Spectral energy transfer functions. All functions are averaged over a period of time in which the convective (baroclinic) dynamics are statistically stationary, as indicated by the time evolution of the vertical Reynolds number $\widetilde{Re}$ (see Supplemental Material at [URL will be inserted by the publisher] for time series of kinetic energy and $\widetilde{Re}$ for representative cases). Magnetic field strength (as characterized by the Chandrasekhar number $\widetilde{Q}$) increases from left to right, with a fixed Rayleigh number of $\widetilde{Ra}=160$. Each plot illustrates the energetic contributions to wavenumber $k$ of the depth-averaged (barotropic) flow, from the other baroclinic and barotropic modes ($T_k$ and $F_k$), from the Lorentz force ($L_k$) and from viscous dissipation ($D_k$). Positive (negative) values indicate energy is being transferred to (from) the $k$-th barotropic mode.}
\label{fig:Transfers}
\end{figure*}

Calculating each of the above functions allows for quantifying the transfer of energy across different spatial scales; the results are shown in Figure \ref{fig:Transfers} for $\Rat=160$. 
In the $\widetilde{Q}=0$ case, as previously documented \cite{rubio_upscale_2014}, there is a net transfer of energy to the largest scales of the barotropic mode due mostly to the non-linear interaction with the baroclinic dynamics ($F_k$), and partly to the interaction between different components of the barotropic flow ($T_k$). Since the sum of these two terms is greater than the dissipation $D_k$, there is a net growth of barotropic kinetic energy at large-scales, leading to the formation of the LSV. For large-scale flows, viscous friction can only become important when the flow speeds become large. Because of this, the formation of the LSV leads to a slow growth of the barotropic kinetic energy with time (see Supplemental Material at [URL will be inserted by the publisher] for time series of the kinetic energy for $\widetilde{Ra}=40$ and $\widetilde{Ra}=160$ cases and different values of $\widetilde{Q}$).  For $\widetilde{Q}=0.1$ the presence of the magnetic field allows for the transfer of energy between baroclinic magnetic energy and barotropic kinetic energy (via $L_k$), which contributes to the dissipation of energy at large-scales. Indeed, $\widetilde{Q}\left<\overline{\textbf{B}}\cdot\nabla j'_z\right> = - \widetilde{Q}\left<\sum_{\textbf{k}}k^{-2} |\overline{\textbf{B}}\cdot\textbf{k}|^2\left<\zeta\right>_{\textbf{k}}\right>$ acts as a dissipative term, proportional to the barotropic component of $\zeta_{\textbf{k}}$. Since $F_k$ is still dominant at larger scales, there is a net growth of barotropic kinetic energy and LSV formation.  The net positive transfer of energy in these cases is due to the temporal averages being calculated over a time-span over which the inverse cascade has not been completely saturated. With time, the dissipation (both viscous and ohmic) grows in magnitude and eventually balances the baroclinic-to-barotropic and the barotropic-to-barotropic transfers, but the dominant wavenumber remains $k=1$. As $\widetilde{Q}$ is increased we find that an inverse cascade (towards scales larger than the injection scale $k=10$) is still present. However, $F_k$ and $T_k$ no longer transport energy to the largest scales (notice the kinetic energy peak at $k=5$ for $\widetilde{Q}=2$ in Figure \ref{fig:Transfers}), and the ohmic dissipation ($L_k$) increases in magnitude to counterbalance $F_k$ and $T_k$. LSVs do not form in such cases, leading to a rapid saturation of the kinetic energy. 

\begin{figure}[tbh]
\centering
{\includegraphics[width=.5\textwidth]{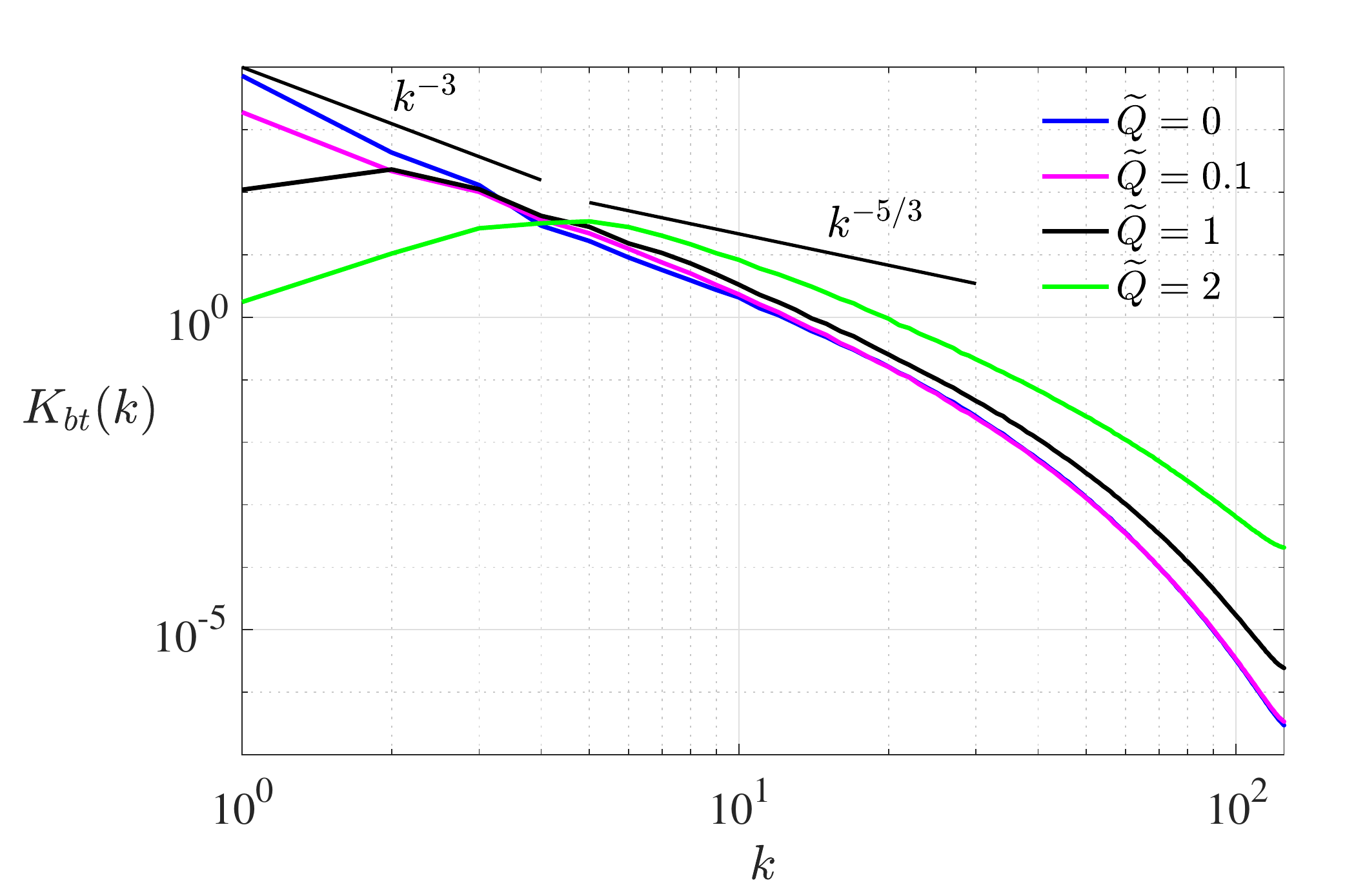}}
\caption{Depth- and time-averaged, barotropic kinetic energy ($K_{bt}$) spectra for a fixed Rayleigh number of $\widetilde{Ra}=160$ and different values of the Chandrasekhar number $\Qt$. Lines of slope $k^{-5/3}$ and $k^{-3}$ are shown for reference. In all cases, energy is injected through convection around $k=10$.}
\label{fig:spectra}
\end{figure}

In Figure \ref{fig:spectra} we show the barotropic kinetic energy spectra for the $\Rat=160$ case. The formation of an LSV for $\widetilde{Q}=0$ and $\widetilde{Q}=0.1$ is evident by the dominance of the box-scale mode. For $\widetilde{Q}=1$ and $\widetilde{Q}=2$ the inverse cascade causes local maxima to be present at $k=2$ and $k=5$, respectively. The $k^{-3}$ slope shown in the plot is expected in the inertial subrange, which is consistent with forward enstrophy cascade \cite{kraichnan1967inertial}, and at the largest scale in presence of large-scale condensates \cite{smith1999transfer,rubio_upscale_2014}. The $k=-5/3$ line is added for reference.

\begin{figure}[tbh]
\centering
\subfloat[]
{\includegraphics[width=.5\textwidth]{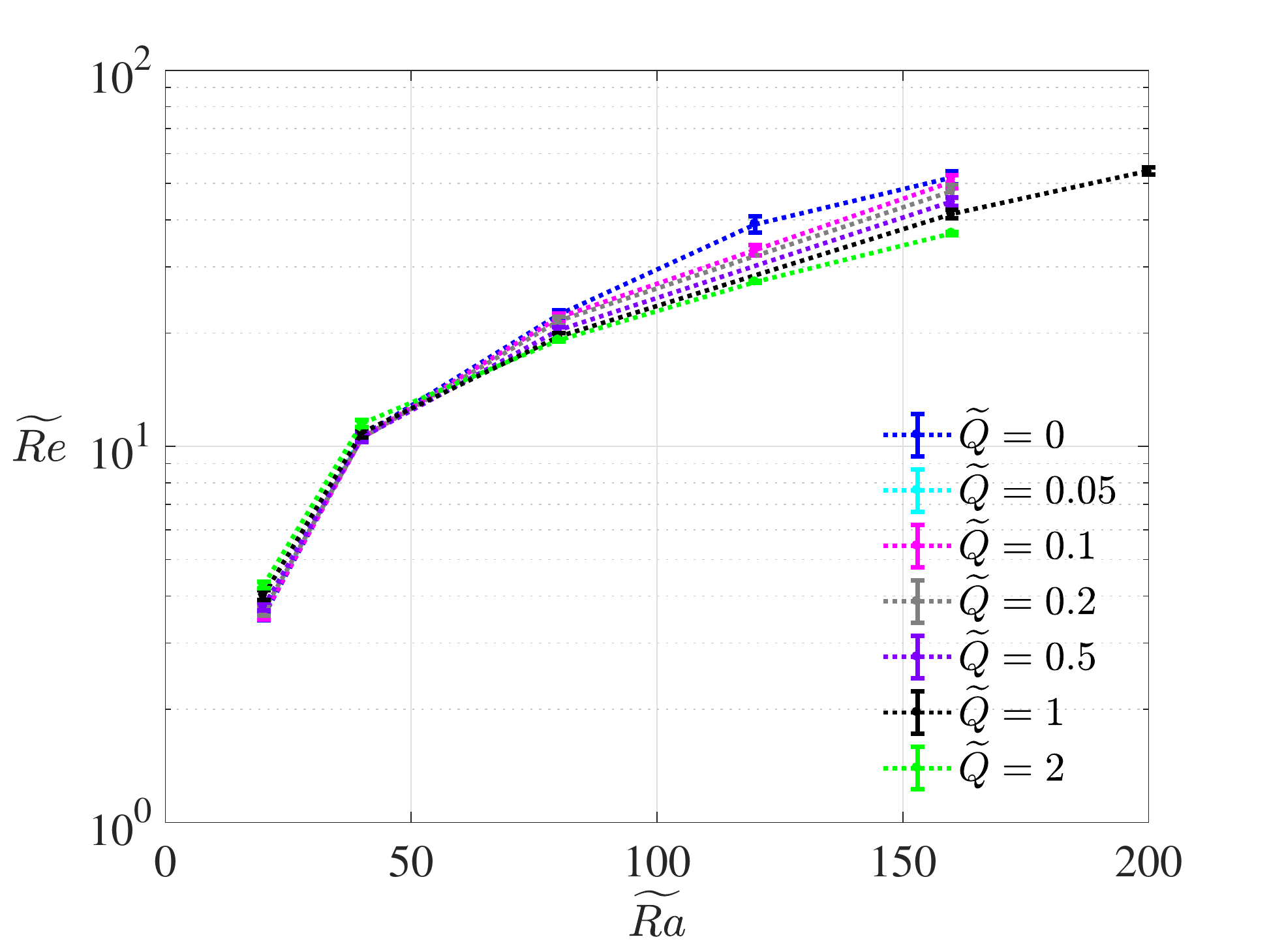}\label{subfig:ReRa}}\\
\subfloat[]
{\includegraphics[width=.5\textwidth]{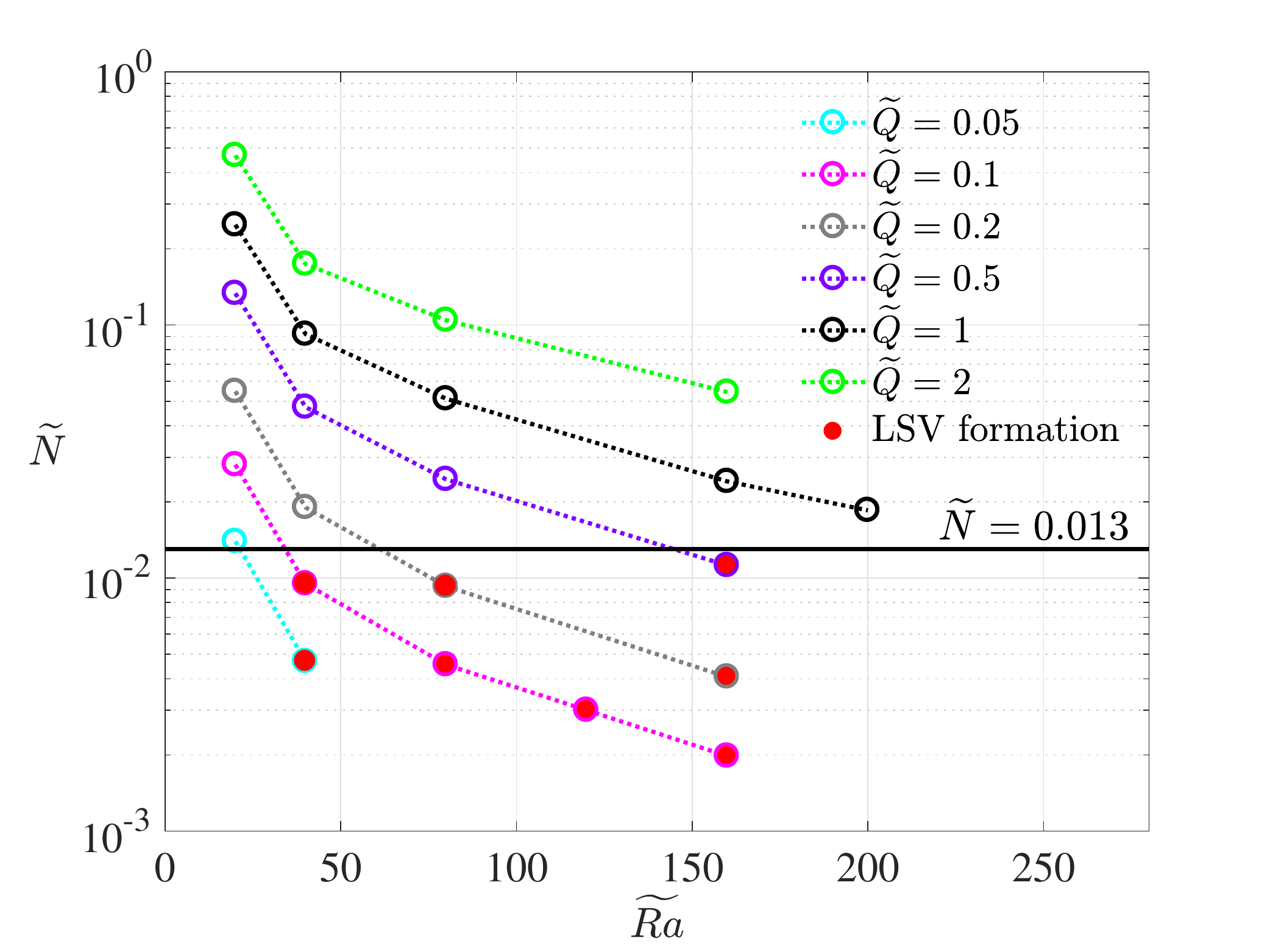}\label{subfig:Nra}}
\caption{(a) Reynolds number $\widetilde{Re}$   and (b) interaction parameter $\widetilde{N}=\widetilde{Q}/\widetilde{Re}$ as a function of $\widetilde{Ra}$ for different values of $\widetilde{Q}$. Values of $\widetilde{Re}$ are time averaged as in Figures \ref{fig:Transfers} and \ref{fig:spectra} and the error bars in (a) represent the fluctuations around the mean value, measured by the standard deviation. Error bars on $\widetilde{N}$ would not be clearly visible on this plot and have been omitted. The filled red circles in (b) indicate large-scale vortex (LSV) formation. The black horizontal line marks the threshold $\widetilde{N}=0.013$ above which no LSV is observed.}
\label{fig:NQ}
\end{figure}

For the values of $\Rat$ and $\Qt$ considered here, Figure \ref{fig:NQ}(a)  shows that the presence of magnetic field does not have an appreciable influence on the convective (vertical) flow speeds, as characterized by the rms small-scale Reynolds number, $\widetilde{Re}$. For $\Rat=160$ the relative difference in $\widetilde{Re}$ is less than $30 \%$ between the $\Qt=0$ and $\Qt=2$ cases. The threshold for LSV formation for $\Qt=0$ is $\widetilde{Re} \gtrsim 10$; all of the magnetic cases with $\Rat \gtrsim 40$ satisfy this hydrodynamic criteria, showing that $\widetilde{Re}$ alone is insufficient for determining when an LSV forms. To better characterize the conditions that favor LSV formation in the presence of magnetic field, we calculate the reduced magnetic interaction parameter $\widetilde{N}$ \cite{cioni2000effect}
\begin{equation}
\widetilde{N}=\frac{\Qt}{\widetilde{Re}} \simeq \frac{|\Qt \, \mBb \cdot \nabla_\perp \bbp|}{|\ubp_\perp\cdot\nabla_\perp\ubp_\perp|} ,
\label{eqn:Ntilde}
\end{equation}
where $\bbp$ and $\ubp$ are the small-scale magnetic and velocity fields, respectively. The interaction parameter is a measure of the relative magnitudes of the Lorentz force and non-linear advection. Figure \ref{fig:NQ}(b) shows that the formation of an LSV is possible for  $\widetilde{N}\lesssim 0.013$. Above this threshold the magnetic field plays a significant role in the dynamics, despite the large $\widetilde{Re}$. The exact threshold value likely depends on the geometry of the mean-field.

\begin{figure*}[tbh]
{\includegraphics[height=7cm]{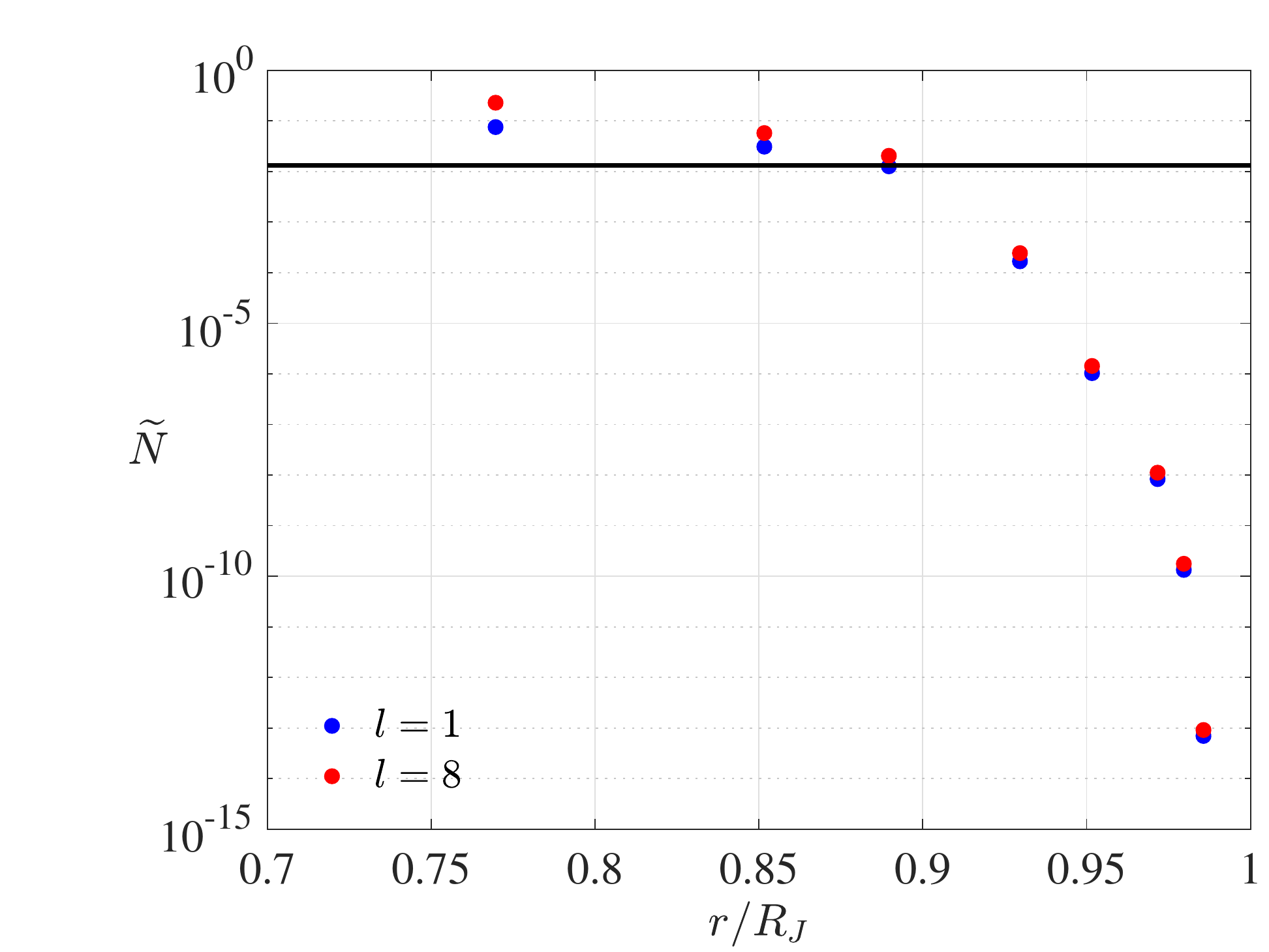}}
\caption{Interaction parameter $\widetilde{N} = \widetilde{Re}^{-1} \widetilde{Q}$ for Jupiter as a function of the non-dimensional radius $r/R_J$, where $R_J$ is the equatorial radius. The blue and red dots are calculated by estimating the large-scale magnetic field intensity to be given by the JRM09 magnetic field model \cite{connerney2018new} truncated at spherical harmonic degrees $l=1$ and $l=8$, respectively. The case $l=1$ corresponds to a dipolar field and for $l>8$ the power in the field components drops below $1 \%$ of the power contained in the dipolar field. Viscosity, electrical conductivity and density are taken from Ref.~\onlinecite{french2012ab}, and $\widetilde{Re}$ is calculated from zonally-averaged meridional flow speeds that are estimated from Cassini spacecraft observations \cite{galperin2014cassini}. The horizontal black line demarcates the $\widetilde{N}=0.013$ threshold.}
\label{fig:jupiter}
\end{figure*}

Our results suggest that it is possible to determine whether LSVs form in natural settings, based on properties that are either directly observable, or inferred from measurements, laboratory experiments and numerical simulations. For instance, ab-initio calculations can be used to constrain the radial variation of density, viscosity and electrical conductivity within Jupiter  \cite{french2012ab}. Magnetic field models of Jupiter, as obtained from the recent Juno spacecraft observations \cite{connerney2018new}, help to estimate $\widetilde{Q}$ in the outermost layers of the planet, where electrical conductivity is small. To estimate $\widetilde{Re}$ we use zonal-mean meridional velocities derived from Cassini spacecraft observations \cite{galperin2014cassini}, which most likely constitute a lower bound  on the convective velocities, and assume they do not change significantly with depth. Together, this data suggests that $\widetilde{N}\ge0.013$ for $r \lesssim 0.87 R_J$, where $r$ is the distance from the center of Jupiter and $R_J$ is the equatorial radius (see Figure \ref{fig:jupiter}). This depth agrees with the location of the dynamo region's upper limit estimated from ab-initio calculations \cite{french2012ab}, from numerical simulation results \cite{duarte2013anelastic} and from the depth of zonal flows based on Juno's gravitational field observations \cite{kong2018origin} (although it is somewhat deeper than the estimated depth reached by deep zonal jets \cite{kaspi2018jupiter,guillot2018suppression,kong2018origin}), suggesting that the observed large-scale vortices \cite{adriani2018clusters} and winds do not penetrate into the dynamo region of Jupiter. For the Earth's outer core we estimate $\widetilde{Re} = 10^3$ and $\widetilde{Q}\approx 3\times10^5$ for accepted values of core flow speed and viscosity \cite{jones20158} and a magnetic field intensity of $3$ mT \cite{gillet2010fast}. These two values give $\widetilde{N}\approx300$, which is well above the threshold of $\widetilde{N}\approx0.013$ identified from Figure \ref{fig:NQ}(b). We conclude that, at the present time, convectively-generated LSVs are likely not present  in the Earth's core.

\section*{Acknowledgements}
This work was supported by the National Science Foundation under grant EAR \#1620649 (SM, MAC and KJ).  This work utilized the RMACC Summit supercomputer, which is supported by the National Science Foundation (awards ACI-1532235 and ACI-1532236), the University of Colorado Boulder, and Colorado State University. The Summit supercomputer is a joint effort of the University of Colorado Boulder and Colorado State University. Volumetric rendering was performed with the visualization software VAPOR. 


%


\bibliography{CU_magnetoconvection.bib}

\clearpage


%
%


\end{document}